\documentclass[prl,aps,onecolumn,preprintnumbers,showpacs]{revtex4}
\usepackage{amssymb,amsfonts,graphicx}

\newcommand{\be}{\begin{equation}}
\newcommand{\ee}{\end{equation}}
\newcommand{\bea}{\begin{eqnarray}}
\newcommand{\eea}{\end{eqnarray}}
\def\id{\mathbb{I}}
\def\bma{\begin{mathletters}}
\def\ema{\end{mathletters}}

\newcommand{\ket}[1]{ | \, #1  \rangle}
\newcommand{\bra}[1]{ \langle #1 \,  |}
\newcommand{\proj}[1]{\ket{#1}\bra{#1}}

\newcommand{\Eins}{\id}
\newcommand{\tr}[1]{\mbox{Tr} \, #1 }
\newcommand{\vis}{\text{v}}
\newcommand{\pr}{\text{P}}
\newcommand{\va}{V_{A}^{(k)}}

\newcommand{\vb}{V_{B}^{(k)}}

\newcommand{\vc}{V_{C}^{(k)}}

\begin{document}

\title{Generation and detection of bound entanglement}

\author{P. Hyllus$^1$, C. Moura Alves$^{2,3}$,
D. Bru\ss$^1$, and C.~Macchiavello$^4$} \affiliation{$^1$Institut
f\"ur Theoretische Physik, Universit\"at Hannover, 30167
Hannover, Germany\\
$^2$Clarendon Laboratory, University of Oxford, Parks Road, Oxford OX1 3PU, UK \\
$^3$Centre for Quantum Computation, DAMTP, University of
Cambridge, Wilberforce Road, Cambridge CB3 0WA, UK
\\
$^4$Dipartimento di Fisica ``A. Volta" and INFM-Unit\'a di Pavia,
Via Bassi 6, 27100 Pavia, Italy}

\date{\today}

\begin{abstract}
We propose a method for the experimental generation of two
different families of bound entangled states of three qubits.
Our method is based on the explicit construction of a
quantum network that produces a purification of the desired
state. We also suggest a route for the experimental
detection of bound entanglement, by employing a witness operator plus
a test of the positivity of the partial transposes.
\end{abstract}
\maketitle

\widetext

\section{1.~Introduction}


Entanglement,
one of the central themes in quantum information processing,
 is well  understood in low-dimensional
systems. In dimensions $2\times2$ and $2\times3$, a
necessary and sufficient condition for entanglement exists:
the partial transposition test \cite{PPT1,PPT2}. However,
the properties of entanglement are much less clear in higher-dimensional
systems, for which only sufficient conditions
for a density matrix to be entangled are known
\cite{higherdim,Terhal,optimization}.
There exist higher-dimensional states that, although entangled, have a
positive partial transpose (PPT). Due to this property,
it is
not possible to distill any entanglement from them with local
operations and classical communication (LOCC).
Undistillable states   are also
called {\em bound
entangled}~\cite{BoundEntanglement}.
 For systems consisting of more than two parties, a state
may be undistillable
even if some of the partial transposes are
non-positive \cite{duer}. Even for bipartite systems,
bound entangled states with non-positive partial transpose probably exist
\cite{undistnpt}.

Apart from the interesting fundamental nature of bound entangled
states, their usefulness
 for quantum information processing has been studied:
bound entangled states can
activate the distillability of one copy of a bipartite
state with non-positive partial transpose~\cite{Activate,Activate2}.
It has also recently been
shown that one can extract a secure key from bound entangled
states~\cite{Key}. In the context
of key creation, results from quantum information
theory, with special use of bound entangled
states, have recently been proven to be fruitful for
insights into open classical information theoretical issues
\cite{boundinfo}.
Various classes of entangled states have been
constructed
theoretically~\cite{be_states}.
However, the topic  of generating bound entanglement in the laboratory
and proving the produced state to be  bound entangled
has not been addressed so far.

How does one generate a certain bound entangled state experimentally?
A solution that is straightforward from a theoretical point of view
is to consider  the spectral decomposition of the state and to compose
a mixed density matrix by creating the eigenvectors with
probabilities that are specified by the
according eigenvalues. However, this
is, in general, a demanding experimental task, as one would need
a source that  can emit various types of product vectors and
entangled vectors with high fidelities
and well-specified probabilities. 
A more satisfactory approach is to
deterministically generate a state that is
the purification of the wanted bound entangled state in some
higher-dimensional Hilbert space. The additional dimensions are
provided by ancilla systems. Then, by tracing out the ancilla
(i.e. experimentally simply ignoring the ancilla part),
one arrives  at the desired bound entangled
state.

In this paper we develop the latter method. Namely, we explicitly
construct quantum networks that generate the two families of bound
entangled states of three qubits introduced in~\cite{abls} and
\cite{duer}. The first family is PPT with respect to any of the subsystems
but nevertheless entangled, while the second family has a
parameter range in which it is NPT only with respect to one subsystem which
is not sufficient for distillation of a singlet between any two of
the parties \cite{duer}. The properties of the latter states have
been used recently in the context of quantum cryptography to show
that so-called bound information exists \cite{boundinfo}. The
networks in both cases act on a six-qubit register that is
initially in state $\ket{000000}$, and from which they generate 
a six-qubit pure state, such that the reduced
density operator $\rho_{\rm bound}$ of the first three qubits is the
desired bound entangled state.

The network for the family \cite{abls} requires only eight
two-qubit gates and one Toffoli gate with three control qubits,
while the network for the family \cite{duer} requires six CNOT
gates, one control-U with two control qubits and one Toffoli gate
with three control qubits. The number of qubits and number of
gates is in foreseeable reach of quantum information technology:
at present, with NMR techniques an order-finding algorithm has
been performed with 5 qubits and 6 control gates \cite{nmr}. In
ion traps, 6 qubits could be  provided, and control gates  and
simple algorithms have been demonstrated \cite{iontraps}.

The second step for the experimental generation of bound entangled states
is to show that the generated states indeed carry bound entanglement.
For the  family of bound entangled states in~\cite{abls} we discuss
this issue explicitly.
The entanglement of the state 
can be proved by using an
entanglement witness 
\cite{optimization}.
We construct an appropriate witness, and provide its local decomposition
 which requires
only four measurements settings. Furthermore, this family of
states has a  PPT with respect to any subsystem. For the
experimental proof of this fact we compare three methods: we
consider the full state estimation of the produced state
$\rho_{\rm bound}$, the more direct spectrum estimation of
$\rho'_{\rm bound}=SPA(\rho_{\rm bound})$, where $SPA$ is the LOCC version of
the structural physical approximation to the partial
transpose~\cite{SPA}, and finally the spectrum estimation of the
partial transpose of $\rho_{\rm bound}$ via the LOCC version of the
network introduced in~\cite{PTnetwork}.

The paper is organized as follows. In Sec.~2 we will introduce
the network that generates the class of bound entangled states
described in~\cite{abls}. In Sec.~3 we construct the entanglement
witness 
that detects entanglement in
the density matrix. 
In Sec.~4 we discuss the three different approaches to
check the positivity of the partial transpositions of the density matrix with
respect to any of the three subsystems.
In Sec.~5, we construct a network that generates the 
family of bound
entangled states of Ref.~\cite{duer} and discuss how the methods
applied in Secs.~3 and 4 could be used 
to experimentally
prove the existence of bound entanglement
in this case.
Finally, in Sec.~6 we conclude
with a summary  of our results.

\section{2.~Generation of bound entangled states}

In this section we explicitly construct the quantum network that
generates the following class of
bound entangled states~\cite{abls}:

\be \rho_{\rm bound}=\frac{1}{N}
\Big(2\proj{GHZ}+a\proj{001}+b\proj{010}+c\proj{011}+\frac{1}{c}\proj{100}+
\frac{1}{b}\proj{101}+\frac{1}{a}\proj{110}\Big), \label{BES} \ee
where $\ket{GHZ}=(\ket{000}+\ket{111})/\sqrt{2}$, the coefficients
fulfill $a,b,c > 0$ and $ab\neq c$, while the normalization reads
$N=2+a+b+c+1/a+1/b+1/c$. This mixed state can be generated
deterministically by a quantum network that uses a register with
three qubits plus three auxiliary qubits, all initialized at
$\ket{0}$, and generates a  pure states of $6$ qubits, such that
the reduced density operator of the three qubits of interest is
$\rho_{\rm bound}$.

\begin{figure}[h]
   \includegraphics[width=10cm]{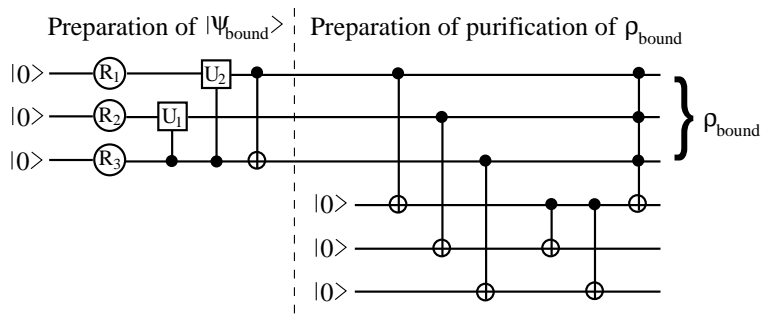}
    \caption{The network for creating the bound entangled
  state given in Eq. (\ref{BES}).}
 \label{fig:circuit}
\end{figure}

The procedure to generate the bound entangled state consists of two
parts: a preparation stage for the first three qubits, and
a purification stage where from the prepared state
and an ancilla state a purification of $\rho_{\rm bound}$ is generated.
In the preparation stage one starts with the three-qubit state
$\ket{000}$, and prepares a three-qubit pure state of the form

 \be \ket{\psi_{\rm bound}}=\frac{1}{\sqrt{N}}
(\ket{000}+\sqrt{a}\ket{001}+\sqrt{b}\ket{010}+\sqrt{c}\ket{011}
+\frac{1}{\sqrt{c}}\ket{100}+\frac{1}{\sqrt{b}}\ket{101}+\frac{1}{\sqrt{a}}
\ket{110}+\ket{111}). \label{psiB} \ee
This is achieved
by applying certain local rotations (LU) on the three qubits, a control-U gate
${\rm CU_{(3,1)}}$
between qubit $3$ and qubit $1$ (qubit $3$ acts as the control
qubit), a control-U gate ${\rm CU_{(3,2)}}$
between qubit $3$ and qubit $2$ (qubit
$3$ acts as the control qubit) and a CNOT gate between qubit
$1$ and qubit $3$ (qubit $1$ acts as the control qubit).
This sequence of gates is illustrated in the left part of
Fig. \ref{fig:circuit}.

The specific form of these gates is given by
\begin{eqnarray}
{\rm LU} &=& N_{1}\left(
\begin{array}{cc}
1 & 1/\sqrt{b} \\
1/\sqrt{b} & -1
\end{array}
\right)_1\otimes N_{2}\left(
\begin{array}{cc}
1 & \sqrt{b} \\
\sqrt{b} & -1
\end{array}
\right)_2 \otimes \left(
\begin{array}{cc}
\alpha & \beta \\
\beta & -\alpha
\end{array}
\right)_3, \\
{\rm CU_{(3,1)}} &=& \Eins_{(1,2)}\otimes \proj{0}_3 +
       N_1 N_3\left(
\begin{array}{cc}
 (\sqrt{a} - {\sqrt{1/bc}}) & (\sqrt{a/b} + \sqrt{1/c})\\
 (\sqrt{a/b} + \sqrt{1/c}) & (-\sqrt{a} + \sqrt{1/bc})
\end{array}
\right)_{1}\otimes\Eins_{2}\otimes\proj{1}_3, \\
{\rm CU_{(3,2)}} &=& \Eins_{(1,2)}\otimes \proj{0}_3 +\Eins_{1}\otimes
N_2 N_4\left(
\begin{array}{cc}
  (1-\sqrt{bc/a}) & (\sqrt{b}+\sqrt{c/a}) \\
  (\sqrt{b}+\sqrt{c/a}) & (-1+\sqrt{bc/a})
\end{array}
\right)_{2}\otimes\proj{1}_3.
\end{eqnarray}
where
$N_1=\sqrt{b/(1+b)},$ $N_2=1/\sqrt{1+b},$
$N_3=\sqrt{c/(1+ac)}$, and  $N_4=\sqrt{a/(a+c)}$.
The coefficients $\alpha$ and $\beta$  depend on $a,b,c$ and must
be chosen such that $\alpha N_1 N_2=\beta N_3 N_4$ and
$\alpha^{2}+\beta^{2}=1$.

It is straightforward to confirm that this
 set of gates is constructed such that it performs the
following sequence of transformations:
\begin{eqnarray*}
    \ket{000} &\stackrel{\rm LU}{\longrightarrow}&
      N_{1}\Big(\ket{0}+\frac{1}{\sqrt{b}}\ket{1}\Big)
      N_{2} \Big(\ket{0}+\sqrt{b}\ket{1}\Big)
      \Big(\alpha\ket{0}+\beta\ket{1}\Big)\\
    &\stackrel{\rm CU_{(3,1)}\cdot CU_{(3,2)}}{\longrightarrow}&
    \frac{1}{\sqrt{N}}\Big[
    \Big(\ket{0}+\frac{1}{\sqrt{b}}\ket{1}\Big)
        \Big(\ket{0}+\sqrt{b}\ket{1}\Big)\ket{0}
    + \Big(\sqrt{a}\ket{0}+\frac{1}{\sqrt{c}}\ket{1}\Big)
    \Big(\ket{0}+\sqrt{\frac{c}{a}}\ket{1}\Big)\ket{1}
    \Big]\\
    &\stackrel{\rm CNOT_{(1,3)}}{\longrightarrow}&
    \ket{\psi_{\rm bound}}.
\end{eqnarray*}

 In the second part of the network one first
applies a sequence of three CNOT
gates between the main and the auxiliary qubits: in this way each term of
$\ket{\psi_{\rm bound}}$ is copied
to the ancilla system. Here, the first, second and third
qubits of the main system act as control qubits, and the first,
second and third ancilla qubits act as target qubits,
respectively:
\bea
 \ket{\psi_{\rm bound}} \ket{000} \stackrel{\rm 3\ CNOTs}
 \longrightarrow \frac{1}{\sqrt{N}}
(\ket{000}\ket{000}&+&\sqrt{a}\ket{001}\ket{001}
+\sqrt{b}\ket{010}\ket{010}+\sqrt{c}\ket{011}\ket{011}
+\frac{1}{\sqrt{c}}\ket{100}\ket{100}\nonumber\\
&+&\frac{1}{\sqrt{c}}\ket{101}\ket{101}
+\frac{1}{\sqrt{a}}\ket{110}\ket{110}+\ket{111}\ket{111}).
\label{psig} \eea
Applying ${\rm CNOT_{(4,5)}}$ and ${\rm CNOT_{(4,6)}}$ then leads
to
\bea
 \longrightarrow & &\frac{1}{\sqrt{N}}
    (\ket{000}\ket{000}+\sqrt{a}\ket{001}\ket{001}
      +\sqrt{b}\ket{010}\ket{010}+\sqrt{c}\ket{011}\ket{011}\nonumber\\
        & & + \frac{1}{\sqrt{c}}\ket{100}\ket{111}
    +\frac{1}{\sqrt{c}}\ket{101}\ket{110}
  +\frac{1}{\sqrt{a}}\ket{110}\ket{101}+\ket{111}\ket{100}). \eea
Finally, one applies a 3-Toffoli gate, where  the three
system qubits are the control qubits and the first auxiliary qubit is
the target. Its action is defined as \cite{NC}

\be \ket{a,b,c}\ket{f}\to \ket{a,b,c}\ket{a\cdot b\cdot c\oplus
f}\;. \ee
The resulting state of the total system is then

\bea \ket{\Psi_{\rm bound}}=\frac{1}{\sqrt{N}} \Big(
(\ket{000}+\ket{111})\ket{000}&+&\sqrt{a}\ket{001}\ket{001}
+\sqrt{b}\ket{010}\ket{010}+\sqrt{c}\ket{011}\ket{011}
+\frac{1}{\sqrt{c}}\ket{100}\ket{111}\nonumber\\
&+&\frac{1}{\sqrt{b}}\ket{101}\ket{110}
+\frac{1}{\sqrt{a}}\ket{110}\ket{101} \Big)\ . \label{psig2} \eea
 Tracing over the three
auxiliary qubits, one obtains that the remaining state of the
three system qubits is of the desired form of Eq. (\ref{BES}):

\bea
 \tr_{\rm aux}(\ket{\Psi_{\rm bound}}\bra{\Psi_{\rm bound}}) = \frac{1}{N}
\Big(2\proj{GHZ} &+& a\proj{001}+b\proj{010}+c\proj{011} \nonumber \\
&+& \frac{1}{c}\proj{100}+
\frac{1}{b}\proj{101}+\frac{1}{a}\proj{110}\Big)= \rho_{\rm bound}.
\label{psig3} \eea
The total quantum network that generates the bound entangled state
$\rho_{\rm bound}$ is shown in Fig. \ref{fig:circuit}.

Note that for the generation of this bound entangled
state  a more general version of the Toffoli
gate can also be applied, namely
$\ket{a,b,c}\ket{f}\to\exp[i\theta(a,b,c)]\ket{a,b,c}\ket{a\cdot b\cdot c\oplus
f}$, because the extra phases cancel when one traces over the ancilla
qubits after the Toffoli gate. This requires less
elementary operations than the Toffoli gate \cite{NC}.
The Toffoli gate with three controls can be decomposed into 13
two-qubit gates \cite{barenco}.
We point out that in this paper we are mainly interested in
providing a network for the generation of bound entanglement with
a small number of  gates, rather than in the optimization of this
network, or the decomposition of the necessary gates into
elementary single and two-qubit gates. The latter issue is
discussed elsewhere in the literature \cite{maslov}.


\section{3.~Construction and decomposition of the entanglement witness}

In this section we will construct and locally decompose an
entanglement witness $W$ that allows to detect the entanglement of
$\rho_{\rm bound}$ with only four local measurements. An entanglement
witness~\cite{Terhal,optimization} is an operator with
non-negative expectation value on separable states, and with
negative expectation value on some entangled states. Thus, if we
construct an appropriate witness for $\rho_{\rm bound}$ and then
measure its expectation value, the experimental result
\begin{eqnarray}
\tr(W\rho_{\rm bound})<0
\end{eqnarray}
indicates that  $\rho_{\rm bound}$ is entangled. Witnesses are
observables with non-local eigenvectors which would be difficult
to measure directly. However, witness operators  can be decomposed
locally~\cite{localwitness} and thus be easily measured in an
experiment.

The state $\rho_{\rm bound}$ that we want to detect has a positive
partial transpose  with respect to every subsystem, but there is
no product vector $\ket{\phi}$ in its range s.t.
$\ket{\phi^{*_X}}$ is in the range of $\rho_{\rm bound}^{T_{X}}$ (here
$X=A,B,C$; the symbol $T_{X}$ denotes partial transposition with
respect to subsystem $X$, and $^{*_X}$ denotes complex conjugation
with respect to subsystem $X$). Such states are called bound
entangled edge states \cite{optimization}. Any PPT entangled state
cannot be detected by
 decomposable witnesses, i.e. witnesses which are of the form
$W=P+Q^{T_{X}}$, where $P$ and $Q$ are positive
operators.
However, there are methods for constructing witnesses that
detect  bound entangled edge states. We will follow the
methods of Ref.~\cite{optimization} for the
construction of our witness \cite{footn}, namely:

\begin{eqnarray}
W = \bar{W}-\epsilon\id,
\label{eqs:witness}
\end{eqnarray}
where
\begin{eqnarray}
\bar{W} = P+Q_{A}^{T_{A}}+Q_{B}^{T_{B}}+Q_{C}^{T_{C}},
\end{eqnarray}
where $P$ denotes the projector onto the kernel of $\rho_{B}$ and $Q_{X}$
is
the projector onto the kernel of $\rho_{B}^{T_{X}}$.
The parameter $\epsilon$ is given by
\begin{eqnarray}
\epsilon&=&\inf_{\ket{e,f,g}}\bra{e,f,g}\bar{W}\ket{e,f,g},
\end{eqnarray}
from which  $\epsilon>0$ follows~\cite{optimization}. We find that
\begin{eqnarray}
     \bar{W}&=&\frac{1}{2}\Big(\proj{000}+\proj{111}\Big)
    +\frac{1}{1+c^{2}}\Big(c^2\proj{100}+\proj{011}\Big)
    +\frac{1}{1+b^{2}}\Big(\proj{010}+b^{2}\proj{101}\Big)
    \nonumber\\
    & &+\frac{1}{1+a^{2}}\Big(\proj{001}+a^{2}\proj{110}\Big)
    -\Big[\frac{1}{2}+\frac{c}{1+c^{2}}+\frac{b}{1+b^{2}}
    +\frac{a}{1+a^{2}}\Big]
    \Big(\ket{000}\bra{111}+\ket{111}\bra{000}\Big).
\label{witness}
\end{eqnarray}
Employing  the Pauli operators $\sigma_z=\proj{0}-\proj{1}$,
$\sigma_x=\ket{0}\bra{1}+\ket{1}\bra{0}$, and
$\sigma_y=-i\ket{0}\bra{1}+i\ket{1}\bra{0}$,
one can use the decomposition \cite{3qubits}
\begin{equation}
    \ket{000}\bra{111}+\ket{111}\bra{000}=
    \frac{1}{4}\Big(
    \sigma_x^{\otimes 3}-\sigma_x\sigma_y\sigma_y-\sigma_y\sigma_x\sigma_y
    -\sigma_y\sigma_y\sigma_x\Big)=
    \frac{1}{2}\Big(
    \sigma_x^{\otimes 3}-\frac{1}{4}(\sigma_x+\sigma_y)^{\otimes 3}
    -\frac{1}{4}(\sigma_x-\sigma_y)^{\otimes 3}\Big).
	\label{decomp}
\end{equation}
For the last expression the local operators
$\sigma_{x}^{\otimes 3}$ and
$\Big((\sigma_{x}\pm\sigma_{y})/\sqrt{2}\Big)^{\otimes 3}$
have to be measured. All the other projectors
in Eq. (\ref{witness}) can be measured
with a single $\sigma_z^{\otimes 3}$ measurement.
Hence, the measurement of the witness requires only 4 measurement
settings. Using the methods of~\cite{3qubits} this
number can be proved to be optimal. On the other hand,
if state tomography is applied to confirm the positivity
of the partial transposes (c.f. chapter IV.),
then all measurements necessary for the witness with the 
first decomposition of Eq.~(\ref{decomp}) are already 
performed there.

The last step on the construction of our witness is the
computation of the value of $\epsilon$. We use the
parametrization
$\ket{e}=\cos{\theta_{e}}\ket{0}
+\exp{i\phi_{e}}\sin{\theta_{e}}\ket{1}$ and accordingly
for $\ket{f}$ and $\ket{g}$. This leads to
\begin{eqnarray}
    \epsilon&=&
    \inf_{\ket{e,f,g}}\Big[ \frac{1}{2}\Big((c_e c_f c_g)^2
            +(s_e s_f s_g)^2\Big)
    +\frac{1}{1+c^{2}}\Big(c^{2}(s_e c_f c_g)^2+(c_e s_f s_g)^2)
   +\frac{1}{1+b^{2}}\Big(( c_e s_f c_g)^2 +b^{2}( s_e c_f s_g)^2\Big)
   \\
   & & +\frac{1}{1+a^{2}}\Big((c_e c_f s_g)^2
                +a^{2}( s_e s_f c_g)^2\Big)
            -\Big[\frac{1}{2}+\frac{c}{1+c^{2}}+\frac{b}{1+b^{2}}
                +\frac{a}{1+a^{2}}\Big]
       \Big(2\cos(\phi_e+\phi_f+\phi_g) c_e c_f c_g
       s_e s_f s_g\Big)
    \Big], \nonumber
\label{epsilon}
\end{eqnarray}
where $c_{e,f,g}\equiv
\cos\theta_{e,f,g}$ and
$s_{e,f,g}\equiv
\sin\theta_{e,f,g}$.
In 
this equation the phases $\phi_e,\phi_f,\phi_g$
appear only in
 the term $\cos(\phi_e+\phi_f+\phi_g)$. Therefore the phases
can be chosen to be equal to zero, using the following argument:
the term
$\Big(2\cos(\phi_e+\phi_f+\phi_g) c_e c_f c_g
       s_e s_f s_g\Big)$ in  
the above equation has to have a positive sign in order to
minimize $\epsilon$. As the coefficients $c_{e,f,g}$ and
$s_{e,f,g}$ occur only quadratically in all other terms, all of
them can be chosen to be positive. Then $\epsilon$ is minimized
for $\phi_e=\phi_f=\phi_g=0$. We are thus left with $6$ real
parameters. If the parameters $a,b,c$ are determined by the
experimental
set-up, 
then the corresponding value of $\epsilon$ can be
obtained numerically by use of a multivariable minimization
routine.

\begin{figure}
 \includegraphics[angle=-90,width=9cm]{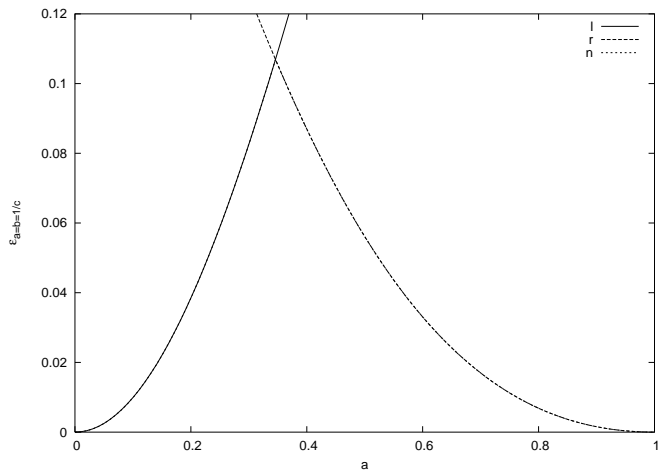}
\caption{Lower bound for ${\epsilon}_{a=b=1/c}$ as a function
of $a$:
The left curve (l) is given by $a^2/(1+a^2)$, while the right 
curve (r) is the analytic minimum of $\epsilon$
for  $\theta_{e}=\theta_{f}=\theta_{g}$.
The maximal lower bound ${\epsilon^{\rm max}}\ge 0.1069$ is obtained for 
$a=0.3460$,
where the two curves meet. The result of the numerical minimization (n) is
 plotted on top of the two analytical curves, and   equals  the
lower branch of them for all $a$.}
\label{fig:eps}
\end{figure}

If the parameters $a,b,c$ can be chosen freely, then it is
advantageous to maximize $\epsilon$ with respect to
$a,b,c$. Making the natural assumption
 that white noise is introduced in the
preparation procedure of the state, {\em i.e.}
$\rho_{p}=p\rho_{\rm B}+\frac{1-p}{8}\Eins,$
the witness will detect entanglement in
the state for $p>1-2\epsilon$. Hence the tolerance of the witness to
the presence of noise is enlarged by maximizing $\epsilon$.
We searched for the maximum in the parameter range
$a=b=1/c\in\, ]0,1[$.
(Remember from the definition of $\rho_{\rm bound}$ in Eq. (\ref{BES}) that
one has to use the open interval here.)
We obtain numerically that for
$a<a_{\rm th}=0.3460$
the minimum is reached at $\epsilon=a^2/(1+a^2)$, {\em i.e.}
when the product state is one of the three possibilities
$\ket{e,f,g}=\ket{011},\ket{101},\ket{110}$.
For $a>a_{\rm th}$ the minimum of $\epsilon$ is obtained when
$\theta_{e}=\theta_{f}=\theta_{g}$. These results are shown
in Fig. \ref{fig:eps}. We find
$\epsilon^{\rm max}_{a=b=1/c}\ge 0.1069$
which is reached for $a_{\rm th}=0.3460$. This is also the
highest value obtained numerically
when $a,b,1/c\in\, ]0,1[$ without the restriction $a=b=1/c$.
For this choice of parameters the state mixed with white noise as described
above is still detected for $p>0.786$, i.e. more than 20\%
of white noise can be tolerated.

\section{4.~Testing the positivity of the partial transpose}

In this section we present three different methods to check the
positivity of the partially transposed density operator
$\rho_{\rm bound}^{T_{X}}$
 with respect to subsystem $X=A,B,C$. One possible
option is to perform the full state estimation of
$\rho_{\rm bound}$~\cite{tomography}, and then to check whether all
the eigenvalues of $\rho_{\rm bound}^{T_X}$ for $X=A,B,C$ are
positive. This method requires the estimation of
$(2\times2\times2)^2-1=63$ independent parameters of the density
operator. This can be achieved by performing $3\times3\times3=27$
measurements on single copies of $\rho_{\rm bound}$, since one can
write any three qubit state as

\begin{eqnarray}
 \rho&=& \frac {1}{8}\sum_{i,j,k=0,x,y,z}\lambda_{i,j,k}\sigma_{i}
\otimes\sigma_{j}
        \otimes\sigma_{k},
\end{eqnarray}
where $\lambda_{i,j,k}=tr(\rho\sigma_{i}\otimes\sigma_{j}
\otimes\sigma_{k})$, and $\sigma_{0}=\id$.
The data from  estimating $\lambda_{l,m,n}$ with $l,m,n=x,y,z$
 can also be used to estimate $\lambda_{0,m,n}$,
$\lambda_{l,0,n}$ and $\lambda_{l,m,0}$. Hence, only local
measurements in the $x,y,z$ directions have to be performed.
One disadvantage of this option is the superfluous estimation of
parameters of the density operator, since we are only interested
in learning about the lowest eigenvalue of the partially
transposed density operator.

Another method for finding out whether $\rho_{\rm bound}^{T_X}>0$ for
$X=A,B,C$ is to start by applying the structural physical
approximation (SPA)~\cite{SPA} to the partial transpose of
$\rho_{\rm bound}$, and then to estimate the lowest eigenvalue of the
resulting density operator. A structural physical approximation is
a completely positive (CP) map, constructed from a positive, but
not CP map, by adding white noise.
The aim in constructing these approximations is to allow the
physical implementation of maps which are useful in entanglement
detection, but are non-physical. In this way one is able to bypass
full state estimation when trying to detect the existence of
entanglement in a given system, since one can  estimate directly
the relevant parameters, e.g the lowest eigenvalue. The
construction of the SPA for a positive, but not completely
positive map $\Lambda$, is as follows:

\begin{eqnarray}
 \widetilde{[\id \otimes \Lambda]}(\rho) =
 \frac{d^4\lambda}{d^4\lambda+1}\frac{I\otimes
 I}{d^2}+(1-\frac{d^4\lambda}{d^4\lambda+1})[\id \otimes
 \Lambda](\rho),
\end{eqnarray}
where $d$ is the dimension of each of the two subsystems on which
$[\id \otimes \Lambda]$ acts, and $\lambda$ is the absolute value
of the most negative eigenvalue obtained when $[(\id \otimes
\id)(\id \otimes \Lambda)]$ acts on the maximally entangled state
$\sum_{i=1}^{d^2} \ket{i}\ket{i}/\sqrt{d^2}$. Each state $\ket{i}$
pertains to a $d^2$-dimensional system, itself composed of two
subsystems of dimension $d$. If one takes $\Lambda$ to be the
transposition map $T$, one finds $\lambda=1/d$. In the two-qubit
case, one obtains that

\begin{eqnarray}
\widetilde{[\id \otimes T]}(\rho) = \frac{2}{9} I\otimes I+
\frac{1}{9} [\id \otimes \Lambda](\rho),
\end{eqnarray}

which can be implemented as

\begin{eqnarray}
 \widetilde{[\id \otimes T]}(\rho) &=&
 \frac{1}{3}\Lambda_1\otimes
 \Lambda_2+\frac{2}{3}\id \otimes
 \sigma_x\sigma_z\Lambda_1\sigma_z\sigma_x,
\end{eqnarray}
where $\Lambda_1(\rho)=1/3\sum_{i=x,y,z}\sigma_i\rho\sigma_i$, and
$\Lambda_2(\rho)=1/4\sum_{i=0,x,y,z}\sigma_i\rho\sigma_i$. Note
that the map $\widetilde{[\id \otimes T]}$ can be implemented
using only LOCC. The extension of this construction to a system of
three, rather than two, qubits is trivial. All we need to consider
is the map

\begin{eqnarray}
 [\id \otimes \widetilde{\id \otimes T}](\rho) &=&
 \frac{1}{3}\id\otimes\Lambda_1\otimes
 \Lambda_2+\frac{2}{3}\id\otimes \id \otimes
 \sigma_x\sigma_z\Lambda_1\sigma_z\sigma_x,
\end{eqnarray}
since the composition of a CP map with identity is still a CP map,
and the construction of $\widetilde{[\id \otimes T]}$ is
independent of the existence of any additional systems. This map
can again be implemented using only LOCC.

Hence, in order to check the positivity of $\rho^{T_X}$ with
$X=A,B,C$, it is enough to implement $\id \otimes \widetilde{\id
\otimes T}$ on $\rho_{\rm bound}$, and then estimate the lowest
eigenvalue of $\rho_{\rm bound}'=[\id \otimes \widetilde{\id \otimes
T}](\rho_{\rm bound})$. The estimation of the lowest eigenvalue of
$\rho'$ can be achieved bypassing full state estimation,
following~\cite{functionals1,functionals2}.

Consider a typical set-up for single qubit interferometry,
conveniently expressed in terms of quantum gates and networks:
Hadamard gate, phase-shift gate, Hadamard gate, and measurement in
the computational basis $\{\ket{0},\ket{1}\}$. We modify the
interferometer by inserting a controlled-$V$ operation between the
Hadamard gates, where $V$ is the swap operator defined as
$V\ket{\phi}_A\ket{\psi}_B=\ket{\psi}_B\ket{\phi}_A$,
$\forall\ket{\phi},\ket{\psi}$. The control is on the qubit and
$V$ acts on the quantum state $\varrho=\rho_A\otimes\rho_B$. The
interaction between the qubit and the environment $\varrho$ via
the controlled-$V$ leads to a modification of the observed
interference pattern by the factor $\vis
e^{i\alpha}=\tr\left[V(\rho_A\otimes
\rho_B)\right]=\tr\left[\rho_A\rho_B\right]$. The generalization
of the swap operation $V$ to the shift operation $V^{(k)}$
(where $V^{(k)}\ket{\phi_1}\ket{\phi_2}...
\ket{\phi_k}=\ket{\phi_k}\ket{\phi_1}...\ket{\phi_{k-1}}$,
$\forall\ket{\phi_i}$, $i=1,...,k$), and the choice of
$\varrho=\rho^{\otimes{k}}$ as the input state, allows us to
estimate multi-copy observables, $\tr[\rho^k]$, of an unknown
state $\rho$~\cite{functionals1}.

\begin{figure}
	\includegraphics[width=10cm]{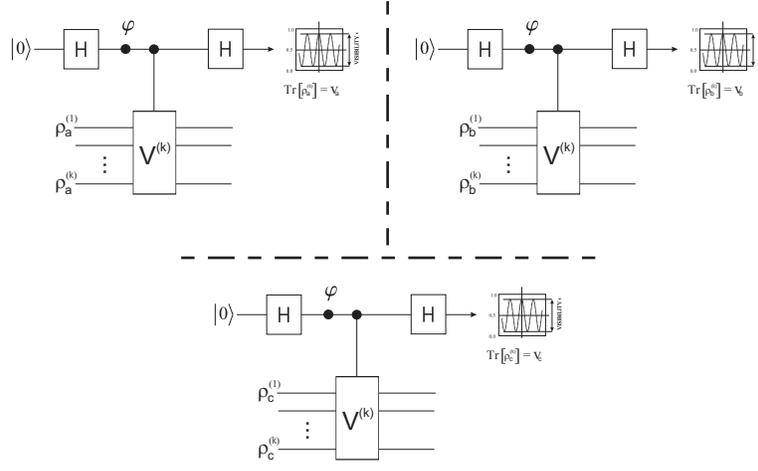}
	\caption{Quantum network that estimates the non-linear functionals
	by LOCC.} \label{swap}
\end{figure}

Let us now extend this method to the LOCC scenario by constructing
three local networks, one for Alice, one for Bob and one for
Charlie, in such a way that the global network is similar to the
network with the controlled-shift. Unfortunately, the global shift
operation $V^{(k)}$ cannot be implemented using only LOCC. Thus,
we will implement it indirectly, using the network depicted
in Fig.~\ref{swap}. 

Alice, Bob and Charlie share a number of
copies of the state $\rho_{ABC}$. They group them respectively
into sets of $k$ elements, and run the local interferometric
network on their respective thirds of the state
$\varrho_{ABC}=\rho_{ABC}^{\otimes k}$. For each run of the
experiment, they record and communicate their results.

The individual interference patterns Alice, Bob and Charlie record
will depend only on their respective reduced density operators.
Alice will observe the visibility $\vis_A=\tr[\rho_{A}^{k}]$, Bob
will observe the visibility $\vis_B=\tr[\rho_{B}^{k}]$ and Charlie
will observe the visibility $\vis_C=\tr[\rho_{C}^{k}]$. However,
if they compare their individual observations, they will be able
to extract information about the global density operator
$\rho_{ABC}$, e.g. about
\begin{equation}
\tr[\varrho_{ABC}^k]=\tr\left[\rho_{ABC}^{\otimes
k}\;\left(\va\otimes\vb\otimes\vc\right)\right].
\end{equation}
This is because Alice, Bob and Charlie can estimate the
probabilities $\pr_{ijl}$ that in the measurement Alice's
interfering qubit is found in state $\ket{i}_A$, Bob's in state
$\ket{j}_B$ and Charlie's in state $\ket{l}_C$, for $i,j,l = 0,1$.
These probabilities can be conveniently expressed as
\begin{small}
\begin{eqnarray}
\pr_{ijl} &=& \frac{1}{8}\tr\Big[\rho_{ABC}^{\otimes k}\; \big(
\id+(-1)^i \va\big) \otimes \big(\id + (-1)^j \vb\big) \otimes
\big(\id + (-1)^l \vc\big)\Big] \nonumber \\
&=& \frac{1}{8}\Big[1 +(-1)^i \tr(\rho_A^k) +(-1)^j \tr(\rho_B^k)
+(-1)^l \tr(\rho_C^k) +(-1)^{i+j} \tr(\rho_{AB}^k) \\
&& +(-1)^{i+l} \tr(\rho_{AC}^k) +(-1)^{j+l} \tr(\rho_{BC}^k)
+(-1)^{i+j+l} \tr(\rho_{ABC}^k) \Big]. \nonumber \label{eq:probs}
\end{eqnarray}
\end{small}
From the latter equality it follows that 
\begin{equation}
	\tr(\rho_{ABC}^k)=\pr_{000}-\pr_{001}-\pr_{100}
		-\pr_{100}+\pr_{011}+\pr_{101}+\pr_{110}-\pr_{111}.
\end{equation}
This is equivalent to 
$\tr(\rho_{ABC}^k)=\langle\sigma_{z}\otimes\sigma_{z}\otimes\sigma_{z}\rangle$,
where $\sigma_{z}=\proj{0}-\proj{1}$.
Given that we are able to directly estimate $\tr[\rho_{ABC}^k]$
for any integer value of $k$, we can estimate the spectrum of
$\rho_{ABC}$ without resorting to a full state tomography.
In our case $\rho_{ABC}\equiv \rho_{\rm bound}$ and we need to
estimate seven parameters, $\tr{\rho_{\rm bound}^j}$ with
$j=2,3,...,8$. Together with $\tr{\rho_{\rm bound}}=1$ they suffice to
determine the eigenvalues of $\rho_{\rm bound}$ \cite{functionals1}.

Even though the SPA option requires the estimation of just seven
parameters of $\rho_{\rm bound}$, it has two potential experimental
difficulties, the first one being the feasibility of implementing
the SPA, and the second one the feasibility of implementing the
quantum networks involving $C-V$ gates.

Finally, we have the option of directly estimating the non-linear
functionals $\tr{[(\rho^{T_X})^k]}$ with $k=1,2,3,...$ and
$X=A,B,C$, following~\cite{PTnetwork}. This scheme is a
modification of the scheme presented in~\cite{functionals1}, and
can be also implemented using only LOCC~\cite{functionals2}. The
main difference between the quantum network of~\cite{PTnetwork},
when compared with~\cite{functionals1}, is that the $C-V^{(k)}$
gates acting on the different subsystems do not  all shift in the
same direction, that is, all but one will shift in the direction

\begin{eqnarray}
V^{(k)}\ket{\alpha_1}\ket{\alpha_2}\ket{\alpha_3}\ldots\ket{\alpha_k}=\ket{\alpha_k}\ket{\alpha_1}\ket{\alpha_2}\ldots\ket{\alpha_{k-1}},
\end{eqnarray}
while the remaining subsystem will shift in the opposite direction

\begin{eqnarray}
V^{-1(k)}\ket{\alpha_1}\ket{\alpha_2}\ket{\alpha_3}\ldots\ket{\alpha_k}=
\ket{\alpha_2}\ket{\alpha_3}\ldots\ket{\alpha_k}\ket{\alpha_1}.
\end{eqnarray}

In fact, the subsystem with respect to which we want to partially
transpose our density operator, will be the subsystem shifted in
the opposite direction. The advantage of this option, when compared 
to the SPA one, is that we do not have to implement any map 
on $\rho_{\rm bound}$ before
estimating the relevant non-linear functionals. Also, the quantum
network used in both schemes has the same level of experimental
difficulty.


\section{5.~Generation of the D\"ur-Cirac-Tarrach states}

Another interesting family of states of three qubits, bound
entagled in a certain parameter range, was introduced in \cite{duer}. 
In this section we will show how to produce it
experimentally, with a method similar to the one described above.
Using the notation from \cite{duer}, this family is given by:
\be
  \rho_{\rm DCT}=\sum_{\sigma=\pm}\lambda_{0}^{\sigma}\proj{\Psi_{0}^{\sigma}}
   +\sum_{k=01,10,11}\lambda_{k}(\proj{\Psi_{k}^{+}}+\proj{\Psi_{k}^{-}}).
  \label{eq:duer}
\ee
Here $\ket{\Psi_{k}^{\pm}}=\frac{1}{\sqrt{2}}(\ket{k_{1}k_{2}0}\pm
\ket{\bar{k}_{1}\bar{k}_{2}1})$, where $k_{1}$ and $k_{2}$ are the
binary digits of $k$,
and $\bar{k}_{i}$ denotes the flipped $k_{i}$.
(Note that the state $\ket{\Psi_0^+}$ in this notation corresponds to
$\ket{GHZ}$ from above.) The
normalization condition reads $\lambda_{0}^{+}+\lambda_{0}^{-}
+2(\lambda_{01}+\lambda_{10}+\lambda_{11})=1$.
With the definitions  $\Delta\equiv\lambda_{0}^{+}-\lambda_{0}^{-}\ge 0$
and
\be
  s_{k}\equiv\left\{
  {
     \textrm{$1$ if $\lambda_{k}<\Delta/2$}
     \atop
     \textrm{$0$ if $\lambda_{k}\ge\Delta/2$}
    }
    \right.
\ee
the following properties of the partial transposes hold \cite{duer}:
\be
  s_{01}=0\Leftrightarrow\rho^{T_{B}}\ge 0,\hspace{0.5cm}
  s_{10}=0\Leftrightarrow\rho^{T_{A}}\ge 0,\hspace{0.5cm}
  s_{11}=0\Leftrightarrow\rho^{T_{C}}\ge 0.
\ee A singlet state between two of the parties can be distilled
iff the partial transposes with respect to the two parties are negative.
For the following choice of the parameters \be
  \lambda_{0}^{+}=\frac{1}{3};\ \ \ \lambda_{0}^{-}=\lambda_{10}=0;\ \
  \lambda_{01}=\lambda_{11}=\frac{1}{6}
  \label{duer_be}
\ee
the corresponding state is inseparable with respect to the
splitting $A-(BC)$ but
separable with respect to the other two splittings.
Hence no singlet can be distilled between any of the parties
and the state is bound entangled.
However, when it is mixed with two states that are obtained by
cyclic permutation of the parties it turns out that the mixture
is inseparable with respect to any partition \cite{activation}.
These properties were used recently to show that
bound information exists and can be activated \cite{boundinfo}.

Let us sketch how the states of Eq.~(\ref{eq:duer}) could be
prepared with our scheme. The density matrix is given by
\be
  \rho_{\rm DCT}=\left(
  \begin{array}{cccccccc}
    \frac{\lambda_{0}^{+}+\lambda_{0}^{-}}{2}&0&0&0&0&0&0&
         \frac{\lambda_{0}^{+}-\lambda_{0}^{-}}{2}\\
    0&\lambda_{11}&0&0&0&0&0&0\\
    0&0&\lambda_{01}&0&0&0&0&0\\
    0&0&0&\lambda_{10}&0&0&0&0\\
    0&0&0&0&\lambda_{10}&0&0&0\\
    0&0&0&0&0&\lambda_{01}&0&0\\
    0&0&0&0&0&0&\lambda_{11}&0\\
    \frac{\lambda_{0}^{+}-\lambda_{0}^{-}}{2} &0&0&0&0&0&0&
          \frac{\lambda_{0}^{+}+\lambda_{0}^{-}}{2}
  \end{array}
  \right).
\ee
We start again with the state $\ket{000}$ and produce the
pure state
\begin{equation}
    \ket{\psi_{\rm DCT}}=
        \frac{\gamma}{\sqrt{2}}(\ket{000}+\ket{100})
        +\sqrt{\lambda_{01}}(\ket{010}+\ket{110})
        +\sqrt{\lambda_{10}}(\ket{011}+\ket{111})
        +\sqrt{\lambda_{11}}(\ket{001}+\ket{101}),
\label{purestate}
\end{equation}
where
$\gamma=\sqrt{\lambda_{0}^{+}+\lambda_{0}^{-}}$.

The state in Eq. (\ref{purestate}) is reached as follows:
 Start by a local rotation and a CNOT gate
\begin{equation}
    \ket{000} \stackrel{\rm LU_{1}}{\longrightarrow}
    \ket{0}(\alpha_{+}\ket{0}+\alpha_{-}\ket{1})\ket{0}
    \stackrel{{\rm CNOT}_{(2,3)}}{\longrightarrow}
    \ket{0}(\alpha_{+}\ket{00}+\alpha_{-}\ket{11})
    \quad{\rm where}\quad
    {\rm LU}_{1} =\Eins\otimes\left(
        \begin{array}{cc}
        \alpha_{+} & \alpha_{-} \\
        \alpha_{-} & -\alpha_{+}
        \end{array}
    \right)\otimes\Eins.
\end{equation}
By proper choice of the coefficients $\alpha_{\pm}$ we can then
reach $\ket{\psi_{\rm DCT}}$ with 3 local unitaries
described below as follows
\begin{eqnarray}
    \stackrel{\rm LU_{2}}{\longrightarrow}
    \ket{0}\big(\gamma\ket{00}
    +\sqrt{2\lambda_{01}}\ket{10}
    +\sqrt{2\lambda_{10}}\ket{11}
    +\sqrt{2\lambda_{11}}\ket{01}\big)
    \stackrel{\rm LU_{3}}{\longrightarrow}\ket{\psi_{\rm DCT}}.
\end{eqnarray}
Hence we have to choose the coefficients and the local unitaries LU$_{2}$
such that
\begin{equation}
    \alpha_{+}\ket{00}+\alpha_{-}\ket{11}
    \stackrel{\rm LU_{2}}{\longrightarrow}
    \alpha_{+}\ket{\phi}\ket{\psi}+\alpha_{-}\ket{\phi^{\perp}}\ket{\psi^{\perp}}
    =\gamma\ket{00}+\sqrt{2\lambda_{01}}\ket{10}
        +\sqrt{2\lambda_{10}}\ket{11}
        +\sqrt{2\lambda_{11}}\ket{01}\big),
\end{equation}
i.e. we have to find the Schmidt decomposition of the state on the RHS
of the last equation. This state has the decomposition
\begin{equation}
    \ket{\varphi}=\sum_{ij}C_{ij}\ket{ij}\quad{\rm with}\quad
    C=\left(
        \begin{array}{cc}
        \gamma & \sqrt{2\lambda_{11}} \\
         \sqrt{2\lambda_{01}} & \sqrt{2\lambda_{10}}
        \end{array}
    \right).
\end{equation}
The Schmidt coefficients are the positive square roots of the
eigenvalues of $C^{T}C$, namely
\begin{equation}
  \alpha_{\pm}^{2}=\frac{1}{2}
  \Big(1\pm\sqrt{1-4[(\gamma^{2}+2\lambda_{01})(2\lambda_{10}+2\lambda_{11})
  -(\gamma\sqrt{2\lambda_{11}}+2\sqrt{\lambda_{01}\lambda_{10}})^{2}]}\Big).
\end{equation}
Then the rotation is given by LU$_{2}=\Eins\otimes V_{2}\otimes U_{2}$,
$U_{2}=(\ket{u+},\ket{u-})$ and $V_{2}=(\ket{v+},\ket{v-})$.
The vectors $\ket{u\pm}$ can be obtained from $(C^{T}C-\alpha_{\pm}\Eins)\ket{u\pm}=0$
and the vectors $\ket{v\pm}$ from $(CC^{T}-\alpha_{\pm}\Eins)\ket{v\pm}=0$.
The last local unitary is given by LU$_{3}=H\otimes\Eins\otimes\Eins$,
where $H=\frac{1}{\sqrt{2}}\left(\begin{array}{cc}1&1\\1&-1\end{array}\right)$,
the Hadamard gate.

\begin{figure}[h]
   \includegraphics[width=10cm]{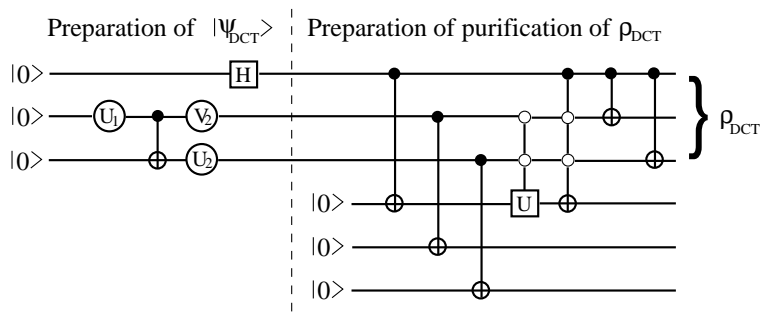}
    \caption{The network for creating the bound entangled
state given in Eq. (\ref{eq:duer}). Open circles for the control bits
indicate that the corresponding gate acts non-trivially
on  the target if the control is
0, rather than 1 as usually (filled circles).
}
 \label{fig:circuit2}
\end{figure}

Now we add again three ancilla qubits in the state $\ket{000}$, and
by using three CNOT gates the first three qubits are copied.
This yields the state
\begin{eqnarray}
    \ket{\psi_{\rm DCT}}\ket{000}&\stackrel{\rm 3\ CNOT's}
{\longrightarrow}&
        \Big(
        \frac{\gamma}{\sqrt{2}}
        (\ket{000}^{\otimes 2}+\ket{100}^{\otimes 2})
        +\sqrt{\lambda_{01}}(\ket{010}^{\otimes 2}+\ket{110}^{\otimes 2})\nonumber\\
        & &+\sqrt{\lambda_{10}}(\ket{011}^{\otimes 2}+\ket{111}^{\otimes 2})
        +\sqrt{\lambda_{11}}(\ket{001}^{\otimes 2}+\ket{101}^{\otimes 2})
    \Big).
\end{eqnarray}
Then we apply the unitary
\begin{equation}
  U=\frac{1}{\gamma}\left(\begin{array}{cc}
    \sqrt{\lambda_{0}^{-}}&\sqrt{\lambda_{0}^{+}}\\
    \sqrt{\lambda_{0}^{+}}&-\sqrt{\lambda_{0}^{-}}
  \end{array}\right)
\end{equation}
on qubit 4 iff the qubits 2 and 3 are in the state $\ket{00}$.
A 2-controlled operation usually acts when both control qubits are
in the state $\ket{1}$, but this can be changed by flipping the
control qubits before and after the 
gate. This operations leads
to the state
\begin{eqnarray}
        & &\frac{1}{\sqrt{2}}\ket{000}
        (\sqrt{\lambda_{0}^{-}}\ket{0}
        +\sqrt{\lambda_{0}^{+}}\ket{1})\ket{00}
        +\frac{1}{\sqrt{2}}\ket{100}
        (\sqrt{\lambda_{0}^{+}}\ket{0}
        -\sqrt{\lambda_{0}^{-}}\ket{1})\ket{00}\\
        & &+\sqrt{\lambda_{01}}(\ket{010}^{\otimes 2}
            +\ket{110}^{\otimes 2})
        +\sqrt{\lambda_{10}}
            (\ket{011}^{\otimes 2}+\ket{111}^{\otimes 2})
        +\sqrt{\lambda_{11}}(\ket{001}^{\otimes 2}
            +\ket{101}^{\otimes 2})
\end{eqnarray}
Then a 3-Toffoli gate flips qubit 4 iff the first three qubits
are in the state $\ket{100}$.
Finally two CNOT gates flip qubits 2 and 3 iff the first qubits'
state is $\ket{1}$. Tracing out the ancilla particles
then yields $\rho_{\rm DCT}$. Summarizing, the procedure is
\begin{eqnarray*}
    \ket{\psi_{\rm DCT}}\ket{000}&
    \stackrel{\rm 3\ CNOT's,^{2}CU,^{3}Toffoli}{\longrightarrow}&
        \sqrt{\frac{\lambda_{0}^{-}}{2}}\big(
            \ket{000}-\ket{100}\big)\ket{000}
        +\sqrt{\frac{\lambda_{0}^{+}}{2}}\big(
                    \ket{000}+\ket{100}\big)\ket{100}
        +\ldots
        \\
    &\stackrel{\rm CNOT_{1,2},\ CNOT_{1,3}}{\longrightarrow}&
    \Big(\sqrt{\lambda_{0}^{-}}\ket{GHZ^{-}}\ket{000}
    +\sqrt{\lambda_{0}^{+}}\ket{GHZ}\ket{100}
    +\sqrt{\lambda_{01}}(\ket{010}^{\otimes 2}+\ket{101}\ket{110})\\
    & &
    +\sqrt{\lambda_{10}}(\ket{011}^{\otimes 2}+\ket{100}\ket{111})
    +\sqrt{\lambda_{11}}(\ket{001}^{\otimes 2}+\ket{110}\ket{101})
    \Big)
\end{eqnarray*}
which leads to
\be
  {\rm Tr}_{4,5,6}\proj{\psi_{\rm DCT}}=\rho_{\rm DCT}.
\ee
The complete network is shown in Fig. \ref{fig:circuit2}.
The existence of bound entanglement for the choice of parameters in
Eq.~(\ref{duer_be}) can be proved by
showing that the state has a PPT with respect to two subsystems,
but not with the third. This can be proved experimentally
by applying the methods of Sec. 4.

Note that the method works for a general choice of the parameters
for which the rank of the density matrix is full.

\section{6.~Conclusions and Acknowledgement}

To summarize,  we have presented a quantum network that generates
bound entangled states of three qubits. Explicitly, we have
studied the production of the two families of bound entangled
states that were introduced in \cite{abls} and \cite{duer}. Note
that our method could be adapted in a straightforward way to the
generation of other types of bound entangled states. As our
networks consists of six qubits and several two-qubit gates, they
go beyond present quantum information processing technology --
however, it seems feasible to realize them in the not too distant
future.

We also discussed
different methods of testing whether the produced states generated by the
network are indeed bound entangled. Namely, we suggested to detect the
entanglement via a suitable witness operator, and to confirm positivity
of the partial transposes by either full state estimation, or spectrum
estimation of the structural physical approximation of the partial
transpose, or direct estimation of some non-linear functionals.

We wish to thank A. Ekert, B. Englert, O. G\"uhne, D. Kaszlikowski, 
and M. \.{Z}ukowski for discussions. We acknowledge support from
Deutsche Forschungsgemeinschaft and the EU (QUPRODIS).
P.H. acknowledges support from the ESF QIT short
scientific visit grant. C.M.A. is supported by the Funda{\c
c}{\~a}o para a Ci{\^e}ncia e Tecnologia (Portugal).

\end{document}